# First *in-vivo* human magnetic particle imaging

P. Vogel [1,4,*], T. Kampf [1,2], M.A. Rückert [1], J. Günther [1,4], T. Reichl [1],
T.A. Bley [3], V.C. Behr [1], P. Gruschwitz [3], V. Hartung [3]

[1] Department of Experimental Physics 5 (Biophysics), Julius-Maximilians University, Würzburg, Germany
[2] Department of Diagnostic and Interventional Neuroradiology, University Hospital Würzburg, Würzburg, Germany
[3] Department of Diagnostic and Interventional Radiology, University Hospital Würzburg, Würzburg, Germany
[4] phase VISION GmbH, Rimpar, Germany

*corresponding author
Patrick Vogel; eMail: Patrick.Vogel@uni-wuerzburg.de

## Abstract

Magnetic particle imaging (MPI) is a tracer-based technique that directly detects the distribution of magnetic iron-oxide nanoparticles with millisecond temporal resolution and no tissue background. Despite extensive preclinical work, in-vivo application of MPI in humans has not previously been reported.

Here, we report the first in-vivo human MPI angiography, visualizing venous perfusion of the upper extremity using a human-scale scanner and clinically approved ferucarbotran. Under identical procedural conditions, we performed X-ray digital subtraction angiography as the clinical gold standard. MPI visualized major superficial and deep veins, including inflow, branching, valve filling, and clearance dynamics in real time with 2 frames per second.

These results establish magnetic particle imaging as a clinically translatable modality for radiation-free vascular imaging in humans and mark the transition of MPI from preclinical research to first clinical application.

## Introduction and purpose

Since the discovery of X-rays in 1895 [1], each new imaging modality has redefined how physicians can see inside the human body. X-ray radiography, Computed Tomography (CT), Magnetic Resonance Imaging (MRI), Ultrasonography (US) and Positron Emission Tomography (PET) each visualize distinct properties and thus different aspects of disease. Each imaging modality provides particular insights and has diagnostic advantages. All carry specific trade-offs such as ionizing radiation, limited temporal resolution or other modality-specific constraints [2]. As a result, these techniques are used in a complementary manner in clinical practice. However, an unmet need remains for radiation-free vascular imaging with high temporal resolution, particularly for image-guided endovascular interventions.

A promising imaging modality is emerging that further extends biomedical imaging: Magnetic Particle Imaging (MPI). First proposed in 2005 [3], MPI directly maps the spatial distribution of magnetic nanoparticles (MNPs) (**see Figure 3**), for example for visualization of the vasculature. Beyond angiography, functionalized MNPs or cell-bound tracers enable the imaging of metabolic and cellular processes [4]. MPI therefore constitutes a tracer-based modality conceptually analogous to Positron Emission Tomography (PET), providing background-free images with millisecond temporal resolution [4,5]. Unlike PET, MPI is free of ionizing radiation and does not apply radionuclides. In contrast to established anatomical imaging modalities such as Computed Tomography (CT) or Magnetic Resonance Imaging (MRI), which often rely on iodinated or gadolinium-based contrast agents, MPI avoids their use. This eliminates associated renal risks and reduces environmental burden [6,7].

In over two decades of preclinical development, MPI has demonstrated its potential for cardiovascular, molecular, and interventional imaging [8-12]. Recent developments now provide all necessary components for clinical translation: human-scale systems capable of high-speed imaging [13,14], clinically approved iron oxide–based tracers with well-characterized safety profiles [15-17] and real-time reconstruction pipelines [18,19].

A key technological development enabling this clinical translation has been the evolution of MPI scanner architectures over the past decade [5]. Early MPI systems were primarily designed for small-animal imaging with limited fields of view. Subsequent advances in selection-field generation and scanner geometry led to the development of scalable concepts such as traveling wave based MPI systems (TWMPI), in which the field-free region (FFR) is dynamically propagated through the imaging volume [20]. This approach enables larger imaging volumes with low hardware requirements [21,22] while maintaining high temporal resolution [23] and system stability [24]. Using this technology, a first pre-clinical setup for interventional purposes was demonstrated [25-27]. Building upon this architecture, a human-scale system, the interventional Magnetic Particle Imaging (iMPI) scanner, has been developed specifically dedicated for real-time vascular imaging in clinical environment [13, 14, 28]. These technological advances now provide the foundation for the first clinical applications of MPI (**see figure 1**).

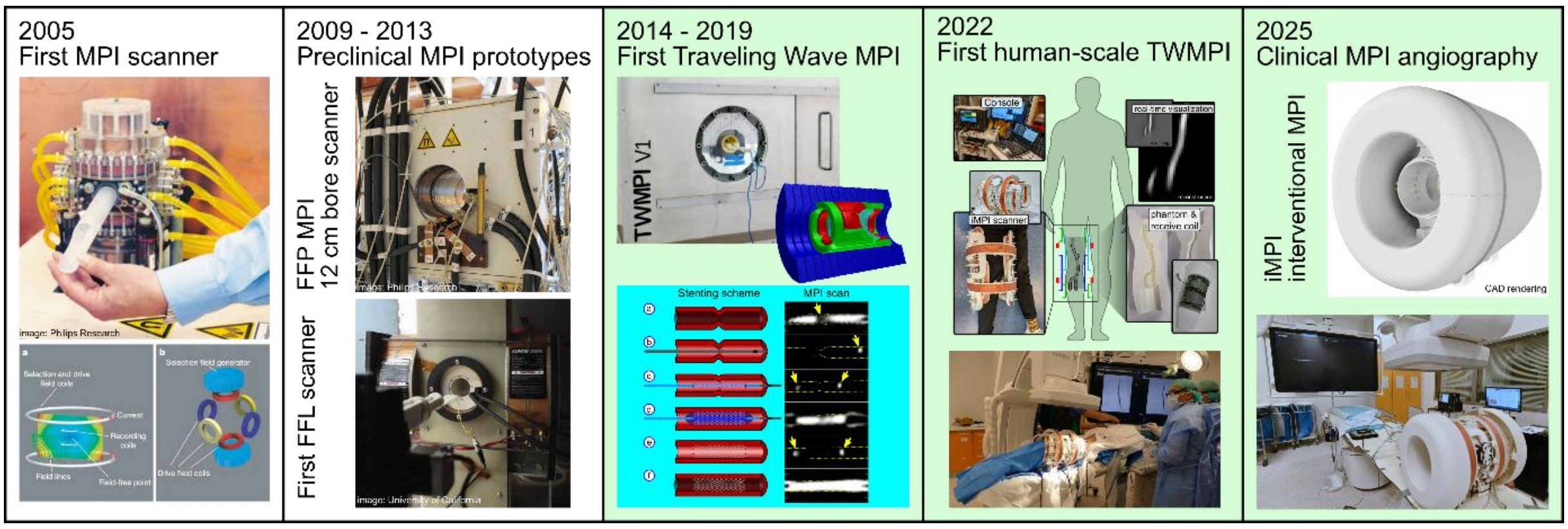

***Figure 1:*** *Evolution of traveling wave magnetic particle imaging (TWMPI) toward clinical MPI angiography (MPA) using the interventional Magnetic Particle Imaging (iMPI) system.*
*(A) Introduction of the first Magnetic Particle Imaging (MPI) scanner, which detects the nonlinear magnetization response of magnetic iron oxide nanoparticles (MNP) in oscillating magnetic fields.*
*(B) Early preclinical MPI systems demonstrated high sensitivity and millisecond temporal resolution but were limited to small-animal imaging with restricted fields of view [29,30].*
*(C) Traveling wave magnetic particle imaging (TWMPI) introduced scalable scanner architectures in which the field-free region is dynamically propagated through the imaging volume [20]. This enables real-time endovascular intervention in a pre-clinical setup [27].*
*(D) These developments enabled human-scale scanners such as the interventional Magnetic Particle Imaging (iMPI) system designed for real-time vascular imaging in an interventional environment [13].*
*(E) In the present study, MPI is translated to the clinical setting with first in-vivo human MPI angiography of the upper extremity venous system.*

In this context, MPI may be utilized for direct angiography of the upper extremity venous system as the first use case. The main indication for upper extremity venous angiography is interventions of arteriovenous fistulas for hemodialysis in patients with chronic kidney disease. First described in 1966, the Brescia-Cimino fistula connects the radial artery and cephalic vein in the forearm and is among the commonly utilized techniques to this day [31]. Although these fistulas are usually established surgically, image-guided, minimally invasive catheter-based endovascular interventions may regularly be required to maintain patency in case of scarring stenoses or thrombosis [32-35]. These interventions may average at one to three endovascular revisions per year, amounting to an estimate of 1 million procedures annually across Europe, North America and Asia [36-38]. These procedures contribute substantially to cumulative radiation exposure for patients and staff, and iodinated contrast agents may be problematic in individuals with residual renal function [39,40].

Here we report the first in-vivo human MPI angiography (MPA), performed under clinical conditions in an angiography lab used for routine patient care by operators trained in endovascular procedures and especially AV-shunt interventions. This radiation-free alternative method is directly compared with the gold-standard digital subtraction angiography (DSA).

# Results

*No procedure-related adverse events occurred. Hemodynamic and clinical monitoring (including 3 channel ecg with automatic arrythmia alert, blood pressure, o2 saturation including breathing frequency) was unremarkable throughout the whole procedure and during post-procedural surveillance. The volunteer did not report any unusual sensations during MPI operation, such as heating, tingling, or twitching.*

DSA demonstrated the expected venous perfusion pattern, including early filling of superficial veins, opacification of the median antebrachial vein, and outflow via the cephalic and basilic veins. Venous valves were clearly visible during the inflow phase. The unphysiologically high pressure while injecting with a pressure bandage in place allowed filling of the deep brachial vein through collaterals, which is a desired effect in clinical imaging to facilitate complete mapping of the venous anatomy.
The iMPI system reproduced all major clinically relevant veins and captured inflow, branching, valve filling, collaterals and clearance dynamics in real time. At 2 frames per second, the temporal resolution was equal to standard DSA settings.
Digital subtraction angiography and MPI were performed sequentially under identical positioning conditions, and the resulting datasets were co-registered offline for direct anatomical comparison.

MPI co-registered to DSA shown in **Figure 2** visualized the same clinically relevant vessels, including the cephalic, basilic, and median cubital veins, as well as valve filling dynamics, collaterals and parts of the deep vein system (brachial vein). Anatomic details of the venous system seen on DSA are presented in **Figure 5** for reference.

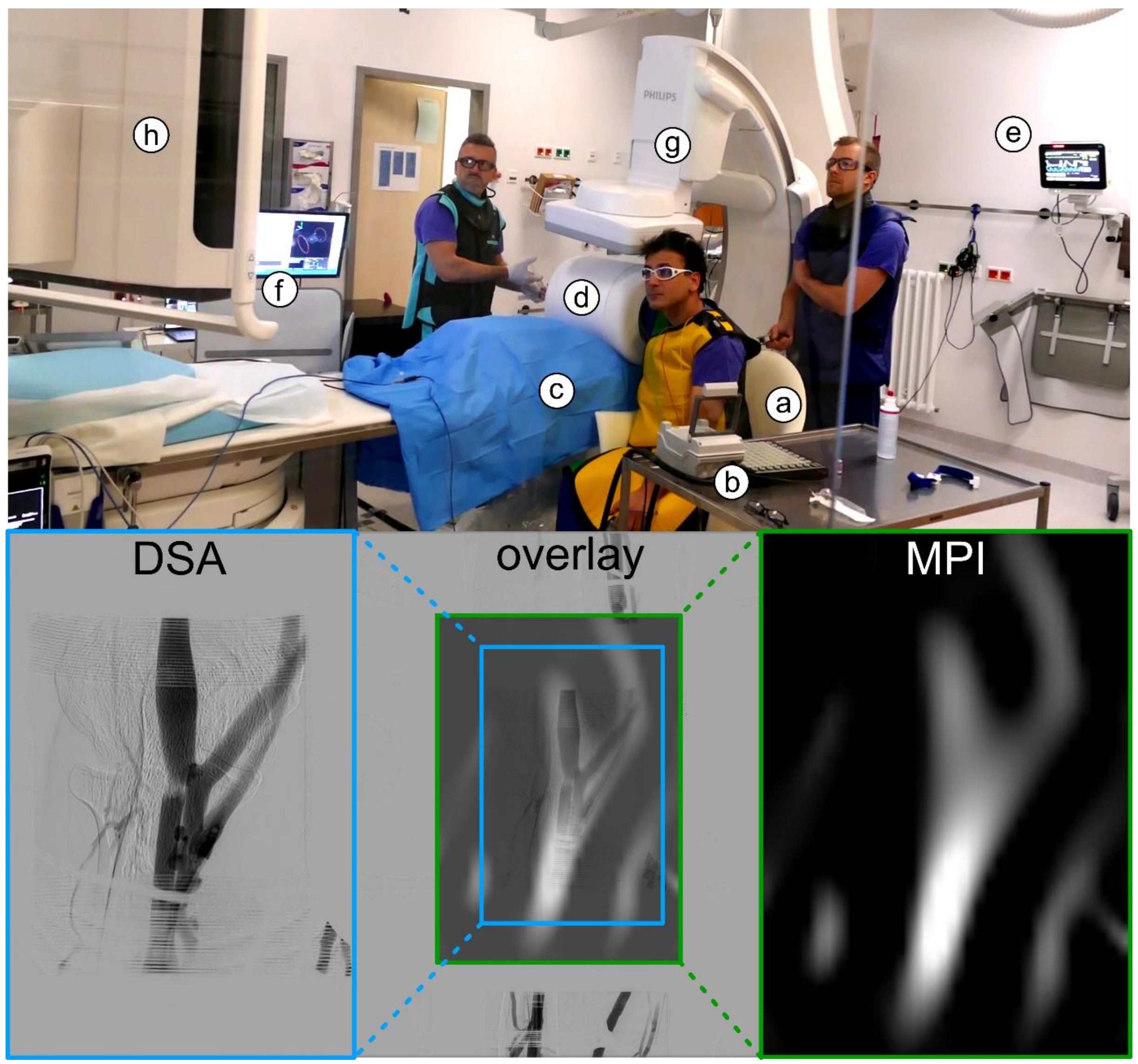

***Figure 2:*** *MPI angiography (MPA) and digital subtraction angiography (DSA).*
***Top****: Overview of the angiography lab during the procedure, showing the operator, volunteer, and independent safety guard.* ***(a)*** *chair for the volunteer,* ***(b)*** *emergency cut-out,* ***(c)*** *patient table,* ***(d)*** *iMPI scanner (interventional magnetic particle imaging system),* ***(e)*** *ECG- and pulse oximeter live monitoring,* ***(f)*** *I/O periphery, amplifier cabinet and live-preview for MPI operation,* ***(g,h)*** *C-arm DSA system (Azurion 7 C20, Philips, Netherlands) with real-time monitoring.*
***Bottom left****: Frame selected from DSA series showing maximum filling of the upper extremity veins.*
***Bottom right****: Corresponding magnetic particle imaging (MPI) angiography frame acquired under identical positioning and injection conditions.* ***Bottom center****: Co-registered overlay of DSA and MPI demonstrating concordant visualization of major superficial and deep veins.*
*The DSA as well as the MPI images are taken directly from the real-time reconstruction series, no postprocessing (****DSA_anim.gif*** *and* ***MPI_anim.gif****).*

# Discussion and Outlook

This study represents the outcome of a structured translational development pathway, spanning methodological groundwork [20], iterative device development of the traveling wave MPI hardware and the interventional iMPI scanner platform [13,18], and validation in cadaveric models [14], designed to ensure technical robustness, procedural safety, and regulatory compliance.

The first in-vivo human application should therefore be viewed not as an isolated proof of concept but as the next step in a structured translational strategy bridging fundamental research and clinical feasibility.

Our findings show that MPI can visualize venous anatomy and flow in real time without ionizing radiation or nephrotoxic contrast, with strong correspondence to the gold standard DSA. By enabling direct and dynamic visualization of blood flow with high temporal fidelity, this advance highlights MPI's potential to support selected interventional and image-guided procedures requiring radiation-free, real-time visualization.

The current limitations in hardware design concern mainly the field of view and scanner size. It would be desirable to increase the field of view of approximately 12 x 8 $cm^2$ to allow a more comprehensive view of pathologies while retaining the lean scanner design capable of integration in a clinical angiography lab, e.g., employing receive arrays to lengthen the field of view as suggested in [21]. Addressing these constraints will be a key objective of future hardware iterations.

The spatial resolution of the current hardware setup of 5 mm [13], which is largely constrained by hardware and safety requirements (SAR), may be sufficient in the context of AV-fistula evaluation. Superficial veins in healthy participants may exceed 10 mm and AV-fistulas usually exceed these diameters by far, even shortly after surgery. In terms of temporal resolution, the iMPI scanner may allow framerates exceeding eight FPS depending on hardware configuration and safety constraints, thereby allowing X-ray fluoroscopy-like temporal resolution without the accompanying dose penalty or loss in spatial resolution.

The utilized tracer dilution of 1:40 ferucarbotran and consequently three image runs (at 20 ml injection volume each; 1.5 ml vial approved for clinical use) resemble improvement over the 1:2 dilution and one image run previously reported with the same scanner [14]. In the context of venous mapping, this may be sufficient for complete mapping of the forearm with the current scanner setup but may not sustain a complete endovascular procedure including intervention, indicating that further optimization of tracer usage and acquisition protocols may enhance procedural coverage.

Another constraint for clinical MPI application may be perceived for patients with metallic implants. Recently, we proved that the observed heating of implants such as stents and orthopedic implants relevant for peripheral vascular interventions is clinically negligible with the current scanner design [28]. However, newer hardware iterations may need confirmation before broader clinical application can be approved by authorities.

More than two decades after its conception, MPI has progressed from preclinical validation to first in-human application. The first in-vivo human demonstration of MPI, performing upper-extremity venous angiography, demonstrated feasibility for clinical vascular imaging. MPI enables radiation-free, real-time imaging with quantitative sensitivity to vascular dynamics, supporting its potential role in selected interventional, cellular and molecular imaging applications alongside established clinical imaging modalities.

In conclusion, this work marks the transition of magnetic particle imaging from preclinical development to first clinical application. By enabling radiation-free, real-time angiography in humans,

MPI has the potential to fundamentally expand the landscape of image-guided vascular and interventional procedures.

# Methods

## MPI basics and MPI scanner

As a tracer-based imaging modality, MPI directly maps the spatial distribution of magnetic iron-oxide nanoparticles [1]. Image contrast arises from the nonlinear magnetization response of MNPs to time-varying magnetic fields, which generates characteristic higher harmonics in the received signal (**Figure 3 (a)**). As biological tissues do not contribute to this signal, MPI provides inherently background-free images with linear quantification of tracer concentration [5].
Spatial encoding is achieved using a strong magnetic field gradient that creates an FFR, in which MNPs are not magnetically saturated and can generate signal (**Figure 3 (b)**). By dynamically moving the FFR along predefined trajectories, the tracer distribution is sampled across the field of view. In current MPI systems, a field-free line (FFL) configuration is employed, offering improved signal-to-noise ratio and faster image acquisition compared to field-free point based (FFP) concepts (**Figure 3 (c)**).
The MPI scanner used in this study is a human-scale traveling wave MPI system designed for endovascular interventions in a clinical setup (iMPI) generating two-dimensional projection images comparable to DSA. Therefore, the FFL is guided across the imaging volume by superimposed selection and drive fields at defined frequencies and amplitudes. The acquired signals are reconstructed in real time using a model-based approach, enabling dynamic visualization of vascular tracer kinetics with high temporal resolution [13,41].

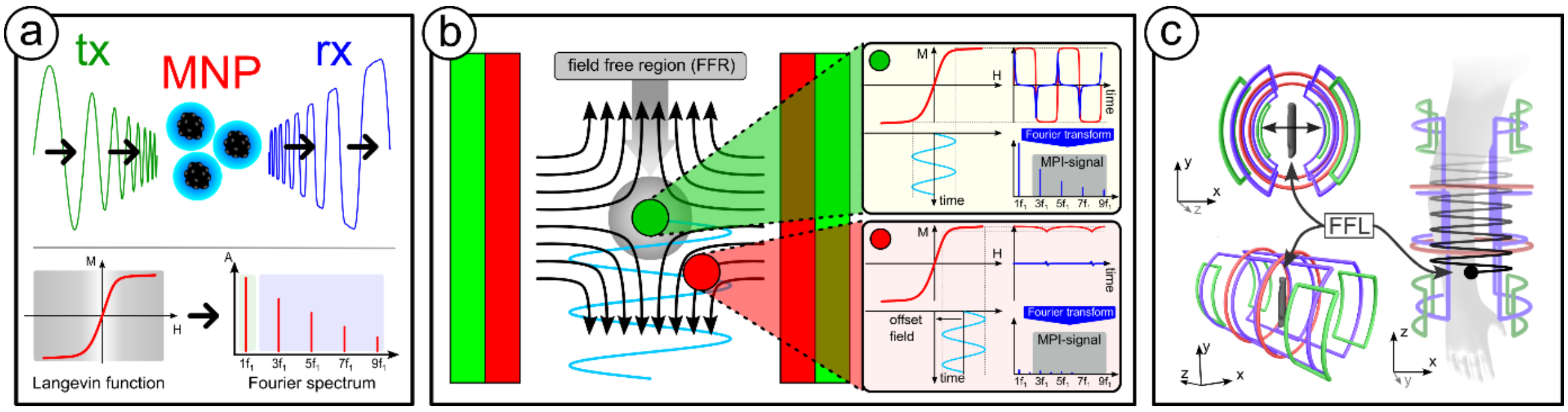

***Figure 3:*** *Magnetic Particle Imaging (MPI) is a tracer-based imaging method that directly maps the spatial distribution of magnetic nanoparticles (MNPs).* ***(a)*** *When an ensemble of MNPs is exposed to an external time-varying magnetic field (tx), its magnetization response follows a nonlinear magnetization curve* M(H)*, the Langevin function, which generates a highly specific signal (rx) in form of higher harmonics in the Fourier spectrum.* ***(b)*** *For imaging, a strong magnetic field gradient is used to select a defined area (field free region – FFR) where the MNPs can generate their specific signal (green region of interest). Otherwise, the offset field suppresses the MNPs signal generation due to saturation effects (red region of interest). By moving the FFR along defined trajectories, the MNP signal can be mapped and visualized.* ***(c)*** *In modern MPI systems, the FFR has the shape of a line (field free line – FFL), which provides several advantages, e.g., a higher signal-to-noise ratio (SNR) and*

*faster signal acquisition. A 2D projection of the field of view (FOV), here a human arm, can be provided by guiding the FFL on a sinusoidal trajectory.*

## Ethics & Safety

The first in-vivo MPA was performed on a healthy volunteer under the direct supervision of a board-certified physician with specialized training in endovascular interventions (European Board of Interventional Radiology, EBIR). The study protocol was reviewed and approved by the institutional review board (IRB; Protocol number: 2025-250-ka), and written informed consent was obtained prior to participation. All procedures were performed under continuous hemodynamic and clinical monitoring following established standard operating procedures for patient care, with ancillary predefined MPI-specific stopping criteria. No adverse events occurred during experiments.

## SAR and PNS

Safety limits defined by the specific absorption rate (SAR) and peripheral nerve stimulation (PNS) constrain the admissible scan parameters in MPI, including magnetic field amplitude, frequency, magnetic field gradient and scan speed. The SAR limit restricts tissue energy deposition to typically 2 $\text{Wkg}^{-1}$ and scales with $f^2B^2$, where $f$ is the main excitation frequency and $B$ the magnetic field amplitude [42].

For the given coil geometry and target gradient of approx. 0.2 T/m, magnetic field amplitudes of about 20 mT are required for guiding the field-free line across the field of view. Operating the selection field at 60 Hz and drive field at 2,620 Hz results in a SAR of approximately 0.1 $\text{Wkg}^{-1}$, well below the safety limit and allowing for potential gradient increases [43]. For drive frequencies below 42 kHz, PNS is the dominant safety constraint. PNS arises from electric fields induced by time-varying magnetic fields and manifests as sensory or motor responses, such as tingling, fasciculation or muscle twitching [44]. Based on the fundamental law of magnetostimulation, the peak-to-peak magnetic field amplitude at the PNS threshold can be estimated from the asymptotic threshold $\Delta B_{\text{min,pp}}$ and the chronaxie time $\tau_c$. Using experimentally determined values for the human arm at 2,620 Hz yields a PNS threshold of approximately 120 mT [43], far above the operating amplitude of 20 mT used in the MPI scanner.

## Experimental Setup

The experiment was prepared in a standard clinical angiography lab at a tertiary care University Hospital (authors' institution) (**Figure 4**). The MPI scanner (**Figure 4 (d)**) was integrated into the angiography environment and positioned on to the patient table, allowing seamless transition between MPI and X-ray–based DSA under identical procedural conditions.

The volunteer was seated next to the angiography table (**Figure 4 (a)**) with the upper extremity placed inside the MPI scanner bore, while continuous hemodynamic monitoring and emergency safety systems were operational at all times. The setup enabled direct comparison of MPI and conventional angiography within a shared field of view and under routine clinical workflow conditions.

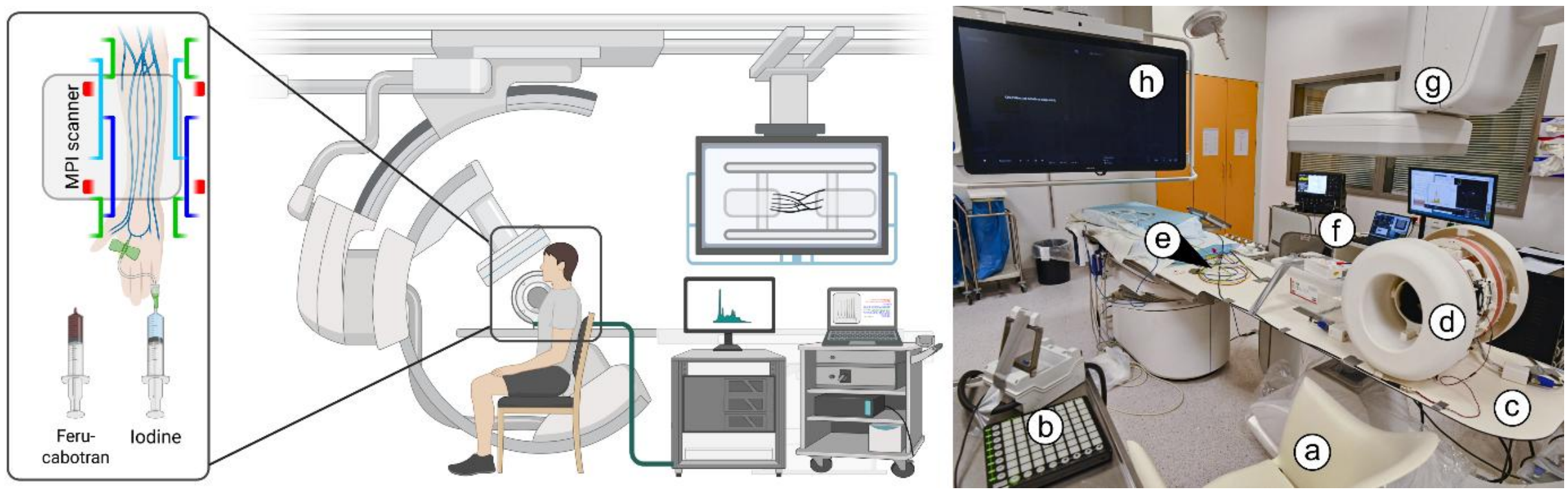

***Figure 4: Left:*** *Schematic of the experimental setup with the volunteer sitting next to the patient table, the arm in the MPI scanner and the X-ray detector unit in place. Large display for angiography visualization and I/O periphery and Amplifier cabinet for MPI operation on the right, magnification of the forearm with cannula for contrast and tracer injection on the left (created in BioRender. Gruschwitz, P.; h23x51b).*
***Right****: On-site setup in the angiography lab with chair for the volunteer* ***(a)****, emergency cut-out* ***(b)****, patient table* ***(c)****, iMPI scanner with opened cover for maintenance* ***(d)****, ECG- and pulse oximeter cables for hemodynamic monitoring* ***(e)****, I/O periphery and amplifier cabinet for MPI operation* ***(f)****, and C-arm DSA system (Azurion 7 C20, Philips, Netherlands)* ***(g)*** *with real-time monitoring* ***(h)****. The operator would stand between the table and MPI periphery* ***(c, f)****. The independent clinical safety guard would stand behind the chair for the volunteer.*

## Endovascular procedure

The experimental setup allows a shared field of view (FOV) of approximately 12×8 cm², which can cover areas such as hand, forearm, elbow and distal half of the upper arm in multiple scans, thereby allowing visualizing of the relevant venous circulation in the context of AV shunt interventions (**Figure 5**). The cubital region, which covers the proximal forearm and elbow was chosen as the region of interest for concomitant DSA and MPI imaging, as selected under fluoroscopic guidance through a dedicated X-ray window of the iMPI scanner. This is reminiscent of AV fistula visualization for Brescia-Cimino fistulas. After establishing a peripheral venous access on the back of the hand and placing a standard 18G cannula, the region of interest was placed in the MPI scanner and/or X-ray detector while aligning FOV under X-ray fluoroscopy.
Following established standard operating procedures for venous mapping in patients, a pressure bandage was applied to transiently reduce venous outflow and enhance contrast filling. Afterwards, 20 ml of contrast agent was injected (50:50 saline and Ultravist 370 (Bayer, Germany)) manually while engaging X-ray digital subtraction angiography to opacify and map the venous vascular anatomy. When filling was complete, the bandage was deflated to allow outflow with contrast clearance promoted by flushing with 20 ml of saline.
In patients, this procedure would facilitate identification of flow-limiting stenoses, thrombosis, aneurysms or collaterals and endovascular procedures may ensue to treat these conditions.

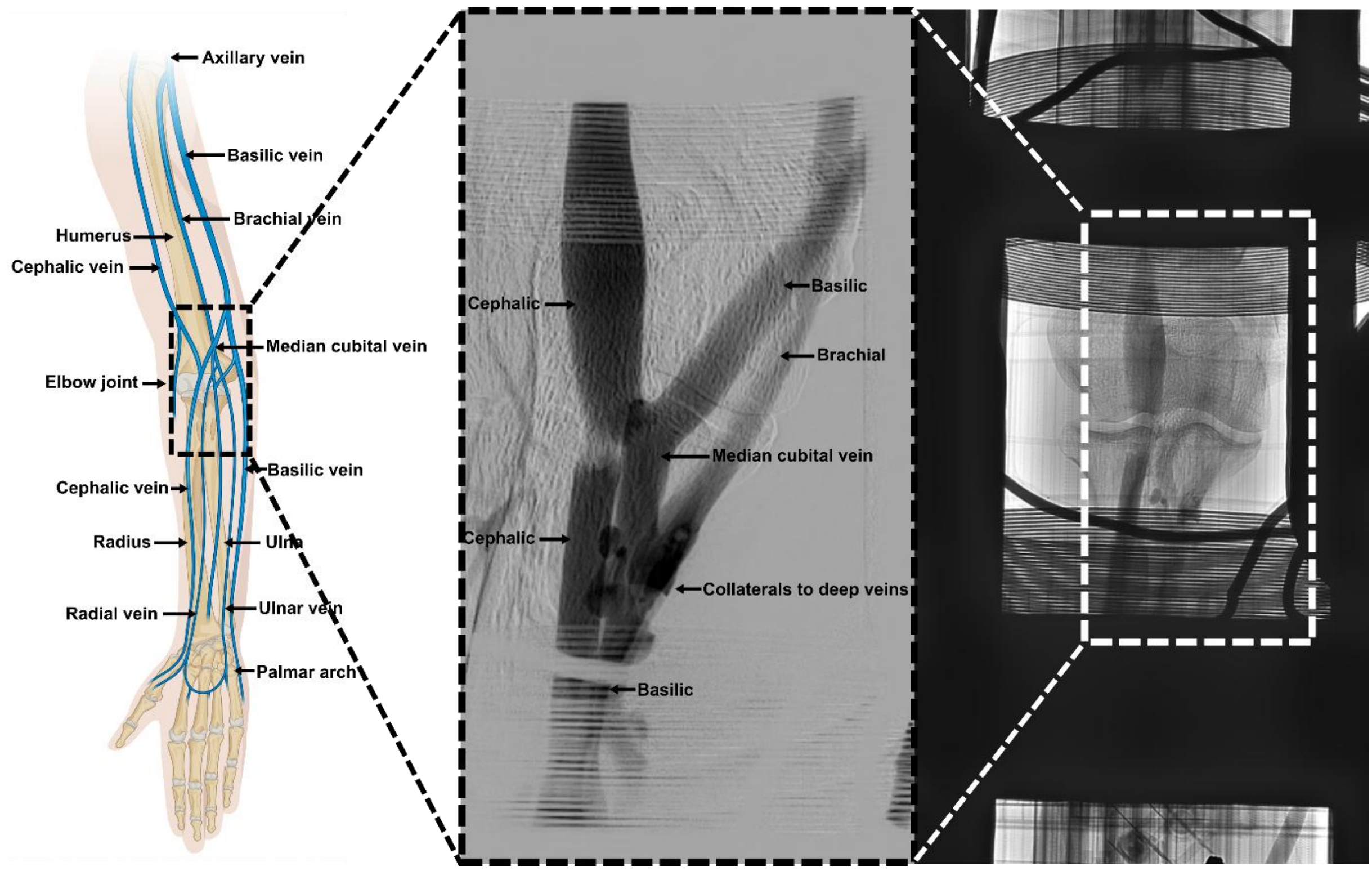

***Figure 5:*** *Digital Subtraction Angiography for mapping of venous Anatomy.*
***Left****: Schematic of the venous anatomy of the upper extremity with superficial veins (basilic and cephalic) and deep veins (radial, ulnar and brachial) which run in parallel to the arteries (not shown). In case of forearm AV-fistula, a forearm vein (i.e. cephalic or median cubital) may enlarge significantly due to direct arterial inflow, thereby facilitating access for hemodialysis (created in BioRender. Gruschwitz, P.; ue896n1).*
***Middle****: Digital subtraction angiography from the shared field of view of the X-ray detector and MPI scanner.*
***Right****: Frame from X-ray fluoroscopy while contrast administration. Shown is the elbow joint seen through the X-ray window of the overlying MPI scanner and coils. Notice the opacification of the basilic, cephalic, median cubital and brachial veins.*

## MPI procedure

With the procedure rehearsed under DSA, a total volume of 20 ml tracer at a dilution of 1:40 (Ferucarbotran (Resotran®, be.imaging, Germany) to saline) was injected with the same procedure, including saline chaser and flushing, while acquiring MPI images at 2 frames per second for 20 seconds. Imaging success was confirmed by examining the reconstructed MPI angiography images, which were available in real-time.
The MPI scanners' operation follows a standardized sequence for creating 2D projection images using a magnetic field gradient of approx. 0.2 T/m at 2.620 Hz drive field frequency and 60 Hz selection field frequency. Real-time image reconstruction was performed with a dedicated MPI framework (Octoview, phase VISION, Germany) using a pre-calculated model-based system matrix [13,41].

## Data availability

The datasets generated and analysed during the current study are available from the corresponding author upon reasonable request.

## Code availability

The reconstruction framework used in this study is available from phase VISION GmbH under a research license upon reasonable request.

## Acknowledgements

We thank Pure Devices GmbH for technical support in the system control of the MPI scanner.

## Funding

The work was supported by the German Research Council (DFG) (grant numbers: VO2288/1-1 [PV], VO-2288/3-1 [PV], BE5293/1-2 [VCB]) and StMWi Bayern: Medical Valley Award (MV-2406-0004).

## Authors contributions

P.V. and V.H. conceived the study.
P.V., T.K., P.G., T.R., J.G., and V.H. designed the experiments.
P.V., T.K., and M.A.R. prepared the hardware and tested the safety components.
V.H., P.G., T.K., T.R., and J.G. performed measurements.
P.V., P.G., and V.H. analyzed the data and prepared the figures.

V.H., P.G., and T.A.B. provided clinical supervision.
P.V. and V.H. wrote the manuscript with input from all authors.
V.C.B. and T.A.B. contributed to manuscript preparation and provided institutional and infrastructural support.
All authors reviewed and approved the manuscript.

## Competing interests

P.V. and J.G. are affiliated with phase VISION GmbH. The remaining authors declare no competing interests.

## Additional information

Supplementary information is available for this paper.
Correspondence and requests for materials should be addressed to P.V.

## Ethics approval

The study was approved by the institutional review board (Protocol 2025-250-ka). Written informed consent was obtained from the participant.
The clinically approved MRI contrast agent Resotran (b.e. Imaging) was off-label used as a tracer.
The scanner used in this investigational context is not approved by the FDA for any purpose.